# The cool brown dwarf Gliese 229 B is a close binary

Jerry W. Xuan[1], A. Mérand[2,†], W. Thompson[3,†], Y. Zhang[1], S. Lacour[2,4], D. Blakely[3,5], D. Mawet[1,6], R. Oppenheimer[7], J. Kammerer[2], K. Batygin[8], A. Sanghi[1], J. Wang[9,10], J.-B. Ruffio[11], M. C. Liu[12], H. Knutson[8], W. Brandner[13], A. Burgasser[11], E. Rickman[14], R. Bowens-Rubin[15], M. Salama[15], W. Balmer[16], S. Blunt[9,10], G. Bourdarot[17], P. Caselli[17], G. Chauvin[18], R. Davies[17], A. Drescher[17], A. Eckart[19,20], F. Eisenhauer[17], M. Fabricius[17], H. Feuchtgruber[17], G. Finger[17], N. M. Förster Schreiber[17], P. Garcia[21,22], R. Genzel[17], S. Gillessen[17], S. Grant[17], M. Hartl[17], F. Haußmann[17], T. Henning[13], S. Hinkley[23], S. F. Hönig[24], M. Horrobin[19], M. Houllé[18], M. Janson[25], P. Kervella[4], Q. Kral[4], L. Kreidberg[13], J.-B. Le Bouquin[26], D. Lutz[17], F. Mang[17], G.-D. Marleau[13,27,28,29], F. Millour[18], N. More[17], M. Nowak[30], T. Ott[17], G. Otten[31], T. Paumard[4], S. Rabien[17], C. Rau[17], D. C. Ribeiro[17], M. Sadun Bordoni[17], J. Sauter[13], J. Shangguan[17], T. T. Shimizu[17], C. Sykes[24], A. Soulain[26], S. Spezzano[17], C. Straubmeier[19], T. Stolker[32], E. Sturm[17], M. Subroweit[19], L. J. Tacconi[17], E. F. van Dishoeck[17], A. Vigan[33], F. Widmann[17], E. Wieprecht[17], T. O. Winterhalder[2] & J. Woillez[2]

[1]*Department of Astronomy, California Institute of Technology, Pasadena, CA 91125, USA*
[2]*European Southern Observatory, Karl-Schwarzschild-Straße 2, 85748 Garching, Germany*
[3]*Department of Physics and Astronomy, University of Victoria, Victoria, BC, V8P 5C2, Canada*
[4]*LESIA, Observatoire de Paris, Université PSL, CNRS, Sorbonne Université, Université Paris Cité, 5 place Jules Janssen, 92195 Meudon, France*
[5]*National Research Council of Canada Herzberg, Victoria, BC, V9E 2E7, Canada*
[6]*Jet Propulsion Laboratory, California Institute of Technology, 4800 Oak Grove Dr.,Pasadena, CA 91109, USA*
[7]*American Museum of Natural History, Department of Astrophysics, Central Park West at 79th Street, New York, NY 10024, USA*
[8]*Division of Geological & Planetary Sciences, California Institute of Technology, Pasadena, CA 91125, USA*
[9]*Center for Interdisciplinary Exploration and Research in Astrophysics (CIERA), Northwestern University, Evanston, IL, USA*
[10]*Department of Physics and Astronomy, Northwestern University, Evanston, IL, USA.*
[11]*Center for Astrophysics and Space Sciences, University of California, San Diego, La Jolla, CA 92093*
[12]*Institute for Astronomy, University of Hawai'i, 2680 Woodlawn Drive, Honolulu, HI 96822, USA*
[13]*Max-Planck-Institut für Astronomie, Königstuhl 17, 69117 Heidelberg, Germany*
[14]*European Space Agency (ESA), ESA Office, Space Telescope Science Institute, 3700 San Martin Drive, Baltimore, MD 21218, USA*
[15]*Department of Astronomy & Astrophysics, University of California, Santa Cruz, CA 95064, USA*
[16]*Department of Physics & Astronomy, Johns Hopkins University, 3400 N. Charles Street, Baltimore, MD 21218, USA*
[17]*Max Planck Institute for extraterrestrial Physics, Giessenbachstraße 1, 85748 Garching, Germany*
[18]*Laboratoire Lagrange, Observatoire de la Côte d'Azur, Université Côte d'Azur, CNRS, Nice, France.*
[19]*1st Institute of Physics, University of Cologne, Zülpicher Straße 77, 50937 Cologne, Germany*
[20]*Max Planck Institute for Radio Astronomy, Auf dem Hügel 69, 53121 Bonn, Germany*
[21]*Universidade do Porto, Faculdade de Engenharia, Rua Dr. Roberto Frias, 4200-465 Porto, Portugal*
[22]*CENTRA, Centro de Astrofísica e Gravitação, IST, Universidade de Lisboa, P-1049-001 Lisboa, Portugal*
[23]*University of Exeter, Physics Building, Stocker Road, Exeter EX4 4QL, United Kingdom*

[24] School of Physics and Astronomy, University of Southampton, Southampton, UK
[25] Department of Astronomy, Stockholm University, AlbaNova University Center, SE-10691, Stockholm, Sweden
[26] Université Grenoble Alpes, CNRS, IPAG, Grenoble, France
[27] Fakultät für Physik, Universität Duisburg-Essen, Lotharstraße 1, 47057 Duisburg, Germany
[28] Institut für Astronomie und Astrophysik, Universität Tübingen, Auf der Morgenstelle 10, 72076 Tübingen
[29] Physikalisches Institut, Universität Bern, Bern, Switzerland.
[30] Institute of Astronomy, University of Cambridge, Madingley Road, Cambridge CB3 0HA, United Kingdom
[31] Academia Sinica, Institute of Astronomy and Astrophysics, 11F Astronomy-Mathematics Building, NTU/AS campus, No. 1, Section 4, Roosevelt Rd., Taipei 10617, Taiwan
[32] Leiden Observatory, Leiden University, P.O. Box 9513, 2300 RA Leiden, The Netherlands
[33] Aix Marseille Univ, CNRS, CNES, LAM, Marseille, France
[†] These authors contributed equally to this work
email: wxuan@caltech.edu

**Owing to their similarities with giant exoplanets, brown dwarf companions of stars provide insights into the fundamental processes of planet formation and evolution. From their orbits, several brown dwarf companions are found to be more massive than theoretical predictions given their luminosities and the ages of their host stars[1–3]. Either the theory is incomplete or these objects are not single entities. For example, they could be two brown dwarfs each with a lower mass and intrinsic luminosity[1,4]. The most problematic example is Gliese 229 B (refs. 5,6), which is at least 2–6 times less luminous than model predictions given its dynamical mass of 71.4 ± 0.6 Jupiter masses ($M_{Jup}$) (ref. 1). We observed Gliese 229 B with the GRAVITY interferometer and, separately, the CRIRES+ spectrograph at the Very Large Telescope. Both sets of observations independently resolve Gliese 229 B into two components, Gliese 229 Ba and Bb, settling the conflict between theory and observations. The two objects have a flux ratio of 0.47 ± 0.03 at a wavelength of 2 μm and masses of 38.1 ± 1.0 and 34.4 ± 1.5 $M_{Jup}$, respectively. They orbit each other every 12.1 days with a semimajor axis of 0.042 astronomical units (AU). The discovery of Gliese 229 BaBb, each only a few times more massive than the most massive planets, and separated by 16 times the Earth–moon distance, raises new questions about the formation and prevalence of tight binary brown dwarfs around stars.**

Gliese 229 B, the first brown dwarf with methane absorption features[5,6], orbits the M1V star Gliese 229 A (0.58 ± 0.01$M_\odot$) with a semi-major axis of 33 AU[1]. The powerful combination of Gaia DR3 and Hipparcos astrometry in addition to decades of imaging and radial velocity (RV) monitoring of the host star enable a precise dynamical mass measurement of 71.4±0.6 $M_{Jup}$ for the companion[1]. The high mass of Gliese 229 B has defied all existing substellar evolutionary models, which predict that a 71.4 $M_{Jup}$ object with an age from 1 to 10 Gyr would have a bolometric luminosity about 2 to 20 times higher than the measured value of $\log(L/L_\odot)$=-5.21±0.05[1,7-9] (see Fig. 3 and Extended Data Fig. 1). In fact, for models that include clouds, 71.4 $M_{Jup}$ is near the hydrogen-burning limit (at solar metallicity) that defines the substellar-stellar boundary[10] (ref. 8: 73.3 $M_{Jup}$; ref. 11: 70.2 $M_{Jup}$). The mass-luminosity discrepancy for Gliese 229 B raises questions about the accuracy of the models, which has serious implications as these models are used to infer masses for most of directly imaged giant planets and brown dwarf companions that lack dynamical masses.

Alternatively, the low luminosity of brown dwarf companions such as Gliese 229 B could be explained if they consist of a spatially unresolved pair of brown dwarfs instead of a single one[1-4]. Other indications of the unusual nature of Gliese 229 B include its near-infrared spectrum, which does not conform to spectral standards, prompting Burgasser et al.[12] to assign it a spectral type of peculiar T7. Despite these anomalies, past observations have unsuccessfully attempted to resolve Gliese 229 B into a binary brown dwarf with adaptive optics imaging[13]. The previous non-detections along with the proximity of the system (5.76 parsec from Gaia[14]), suggest that a putative binary would have a tight separation of <0.2 AU or a small mass ratio[1]. However, known brown dwarf binaries show a strong preference for equal mass ratios and a separation distribution peaking between approximately 1–3 AU (refs. 15,16).

We observed Gliese 229 B on five nights using the Very Large Telescope Interferometer (VLTI) in GRAVITY Wide mode[17] with the Unit Telescopes of the European Southern Observatory (ESO) at Cerro Paranal, Chile. The observations were performed in the K band (1.95–2.45 μm). We extracted closure phases from the GRAVITY data (see Methods), in which a non-zero closure phase indicates a departure from central symmetry, for example, a binary source. As part of the same program, we observed Gliese 229 B with The CRyogenic InfraRed Echelle Spectrograph Upgrade Project (CRIRES+)[18] on UT3 of the Very Large Telescope in the H band (1.5–1.75 μm) on seven different nights to monitor its RV. The CRIRES+ spectra have a resolving power ($\lambda/\Delta\lambda$) of ≈100,000 and were extracted as described in Methods.

We find strongly non-zero closure phases in the first epoch of GRAVITY observations (Extended Data Fig. 2) that are consistent with a binary source. The subsequent GRAVITY epochs confirm the detection and provide evidence of orbital motion between the two components (Fig. 1). With the first epoch alone, the null hypothesis that Gliese 229 B is a single source (i.e., all closure phases should be zero) leads to a reduced $\chi^2$ of 55 (288 degrees of freedom). Carrying out a grid search for the companion as described in ref. 19, we find a secondary brown dwarf located ≈5 mas south of the brighter, primary brown dwarf, with a secondary-to-primary flux ratio of ≈0.5. The binary fit has a much lower reduced $\chi^2$ of 1.27. In the binary fit, we also account for linear motion of the companion over the 2.5-h observing window. We find that the companion moves in a direction nearly perpendicular to the vector between itself and the brighter brown dwarf at a rate of $4.6^{+1.5}_{-0.8}$ mas/day (Fig. 1c), consistent with the expected motion of ≈4.6 mas from a circular, face-on orbit for a total mass of 71 $M_{Jup}$.

Contemporaneous CRIRES+ monitoring independently confirms that Gliese 229 B is a binary brown dwarf. Initially, we cross-correlated the CRIRES+ spectra of Gliese 229 B with a Sonora Elf Owl atmospheric model[20] assuming $T_{eff}$=900 K, log($g$)=5.0 (ref. 21). The cross-correlation functions (CCFs) displayed time-varying line locations and shapes consistent with the partially resolved spectra of two brown dwarfs orbiting each other (Extended Data Fig. 3,4). Therefore, we fit the CRIRES+ spectra as emission from two brown dwarfs and account for a small amount of starlight leakage into the slit using observations of Gliese 229 A (Fig. 2a, see Methods). Based on the GRAVITY measured flux ratio, we started with atmospheric models with $T_{eff}$=850 K and $\log(g)$=5 for the primary brown dwarf and $T_{eff}$=750 K and $\log(g)$=5 for the secondary (see Methods) to extract the RV of each brown dwarf. For each CRIRES+ epoch, alternative fits of the spectra with a single-component brown dwarf model are disfavored with statistical significance ≳ 20σ. The extracted RVs display unambiguous signs of a spectroscopic binary (Fig. 2b).

We combine the CRIRES+ and GRAVITY data to characterize the orbit of the binary brown dwarf. The orbit fits are performed with PMOIRED[22] and Octofitter[23], as described in Methods. The data are well fit by the model with a reduced $\chi^2$ of 2.2 (513 degrees of freedom) and slightly broad but symmetrical closure phase residuals, with the model accounting for all closure phase features. The GRAVITY K band flux ratio is constrained by the joint fit to 0.47 ± 0.03. We derive an orbital period of 12.134 ± 0.003 days, corresponding to a semi-major axis of 0.0424 ± 0.0004 AU, or about 90 Jupiter radii. The ratio of the RV semi-amplitudes directly constrains the mass ratio (q) to $0.90^{+0.06}_{-0.02}$. From the binary brown dwarf's orbit, we independently measure a total mass of 72.5 ± 1.3 $M_{Jup}$, which is consistent with the mass derived by ref. 1 from the orbit of the unresolved Gliese 229 B around Gliese 229 A. We measure component masses of 38.1 ± 1.0 $M_{Jup}$ and 34.4 ± 1.5 $M_{Jup}$, an eccentricity of 0.234 ± 0.004, and inclination of 31.4 ± 0.3°. The eccentricity of Gliese 229 Bab is typical compared to the eccentricity distribution of field brown dwarf binaries[24]. We note that the outer orbit of Gliese 229 Bab around Gliese 229 A is highly eccentric (e≈0.85) and viewed nearly face-on[1]. The binary brown dwarf's orbit is moderately misaligned relative to the outer orbit by $37^{+7}_{-10}$°. Additionally, the host star's spin orientation is viewed nearly edge-on[25] and therefore misaligned relative to both inner and outer orbits.

To make the astrometric and spectroscopic observations fully self-consistent with the atmosphere models, we interpolate the ATMO 2020 substellar evolutionary model[9,26] to search for component masses and ages that simultaneously reproduce the GRAVITY K band flux ratio and bolometric luminosity (see Methods). Adopting a prior on the total mass of 72.5 ±1.3 $M_{Jup}$, we find that a binary brown dwarf with mass ratio of 0.87 ± 0.03 and age of 2.45 ± 0.20 Gyr matches the models well. This mass ratio is consistent at the 1σ level with the value derived from the orbit fit. From ATMO 2020, the primary component is estimated to have $T_{eff}$=860 ± 20 K, $\log(g)$=5.11 ± 0.01 dex, and $\log(L/L_\odot)$=-5.41 ± 0.04, while the secondary component has $T_{eff}$=770 ± 20 K, $\log(g)$=5.03±0.01 dex, and $\log(L/L_\odot)$=-5.58 ± 0.04. Our inferred age agrees with the value of ≈2-6 Gyr estimated for the host star[13]. Therefore, our detection of the binary and measurements of its properties bring the system into much better alignment with substellar evolutionary models, as shown in Fig. 3.

Although the near-unity mass ratio between Gliese 229 Ba and Bb fits with previous brown dwarf binaries[16], the semi-major axis of approximately 0.042 AU makes it the tightest brown dwarf binary in a triple system (Extended Data Fig. 5). Among brown dwarf binaries orbiting stars, the next closest binaries have semi-major axis values more than an order of magnitude larger at about 0.9 AU (for example, Gliese 569Bab (ref. 24)). A few isolated ultracool dwarf binaries with component masses between 0.08-0.09 $M_\odot$ have smaller separations[27,28], but among unambiguous brown dwarf binaries, only 2MASS J0535−0546AB and SPEC J1510−2818AB have comparable separations of 0.04 and 0.06 AU, respectively[29,30]. The formation mechanism of brown dwarf binaries around stars remains an open question, and both observations and simulations are highly incomplete for brown dwarf binaries with separations <1 AU (ref. 15). Opacity-limited fragmentation restricts the primordial separations of objects to distances >10 AU (ref. 31), implying that significant dynamical and dissipative processes are required to form tight brown dwarf binary systems[32]. Although the exact processes for dissipation is unclear, tidal interactions between the gaseous envelopes or accretion disks around the forming objects are likely important[33,34]. For binary brown dwarfs orbiting stars, fragmentation of a massive circumstellar disk is another potential formation route, where two proto-brown dwarfs fragment in the disk and become bound in a close encounter[33]. Ultimately, any formation mechanism would need to account for the highly eccentric outer orbit of Gliese 229 A-Bab and the misalignments between the inner orbit, outer orbit, and host star spin axis.

Thirty years after its discovery, Gliese 229 B continues to teach us about substellar objects. The discovery of Gliese 229 BaBb provides a potential resolution to the mass-luminosity tension for brown dwarf companions and suggests that other unusually massive brown dwarfs, such as HD 4113 C (ref. 2), could be unresolved substellar binaries as well. Future efforts to resolve other anomalous brown dwarf companions into binaries are essential for rigorously testing substellar evolutionary models, which are routinely used to interpret observations of giant planets. Although known brown dwarf binaries have separations peaking between 1-3 AU (ref. 16), Gliese 229 Bab demonstrates the existence of binary substellar companions to stars with separations well below 1 AU. The 12 day orbital period of Gliese 229 Bab places the two brown dwarfs deep within the Hill sphere of each other, suggesting a formation pathway that involves significant energy dissipation. A major goal of exoplanet studies in the next decade is the search for exomoons and binary planets. Among isolated brown dwarf binaries, there are already examples of systems where both components have masses in the planetary regime[35,36] (below 13 $M_{Jup}$), in addition to several systems with ≈4-13 $M_{Jup}$ companions orbiting low-mass brown dwarfs[37-39]. It is unclear how common binary planets or exomoons are around stars. With further improvements in sensitivity, the combination of interferometry, high-resolution spectroscopy, and transit photometry is poised to unveil new discoveries and provide insights into these questions.

**Table 1 | Orbital and physical parameters of Gliese 229 BaBb**

| | Confidence Interval (frequentist analysis): Median and 16-84% percentiles | Credible Interval (Bayesian analysis) Median and 16-84% |
|---|---|---|
| **Orbital period (day)** | 12.134±0.003 | 12.137±0.001 |
| **Semimajor axis (au)** | 0.0424±0.0004 | 0.0422±0.0001 |
| **Eccentricity** | 0.234±0.004 | 0.234±0.002 |
| **Argument of periastron (°)** | 180.7±1.2 | 182.8±0.9 |
| **Inclination (°)** | 31.4±0.3 | 31.1±0.4 |
| **Longitude of ascending node (°)** | 213±2 | 210.3±1.2 |
| **Time of periastron (MJD)** | 60377.88±0.04 | 60377.85 ± 0.02 |
| **Mass ratio ($M_2/M_1$)** | $0.90^{+0.06}_{-0.02}$ | $0.91^{+0.06}_{-0.05}$ |
| **Mass of Ba, $M_1$ ($M_{Jup}$)** | 38.1±1.0 | 37±1 |
| **Mass of Bb, $M_2$ ($M_{Jup}$)** | 34.4±1.5 | 34±1 |
| **Flux ratio, $f_2 / f_1$ (2.0 µm)** | 0.47±0.03 | 0.53±0.02 |
| **$\gamma_{RV}$ (km/s)** | 0.4±0.2 | 0.46±0.20 |
| **Total mass of B ($M_{Jup}$)** | 72.5±1.3 | 71.3±0.5 |

The argument of periastron refers to the primary brown dwarf, Gliese 229 Ba.

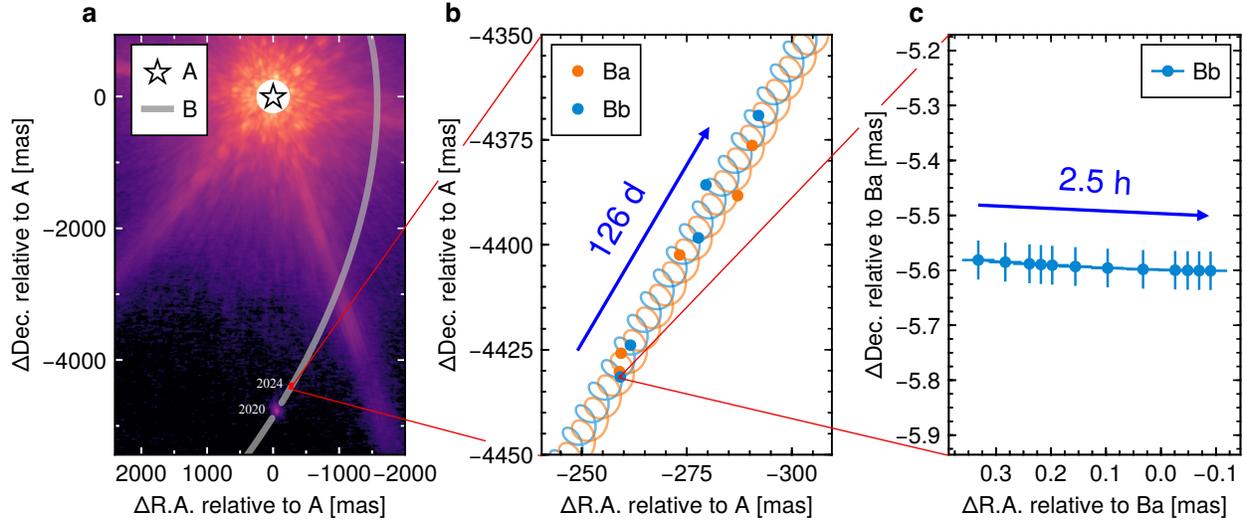

**Fig. 1 | The detection and astrometric orbit of Gliese 229 BaBb.** (a) A Keck/NIRC2 $K_s$ band image of the Gliese 229 system taken on Oct 18, 2021. The binary brown dwarf is unresolved given Keck's resolution of 45 milliarcsecond. The gray line indicates the best estimate of the outer orbit of Gliese 229 BaBb around A[1]. (b) A zoom-in for the maximum a-posteriori binary brown dwarf orbit from the GRAVITY and CRIRES+ joint fit, where measured positions of Gliese 229 Ba and Bb in each GRAVITY epoch are shown as orange and blue points, respectively. The average uncertainty on the derived relative position between Bb and Ba is between 0.01 and 0.05 milliarcseconds. Note that GRAVITY and CRIRES+ only measure the differential positions between Ba and Bb, so the length and direction of the spiral pattern are derived from the maximum a-posteriori draw of the outer orbit (gray line in panel a). (c) The motion of Gliese 229 Bb relative to Gliese 229 Ba during the 2.5 h observing window of the first night of GRAVITY observations.

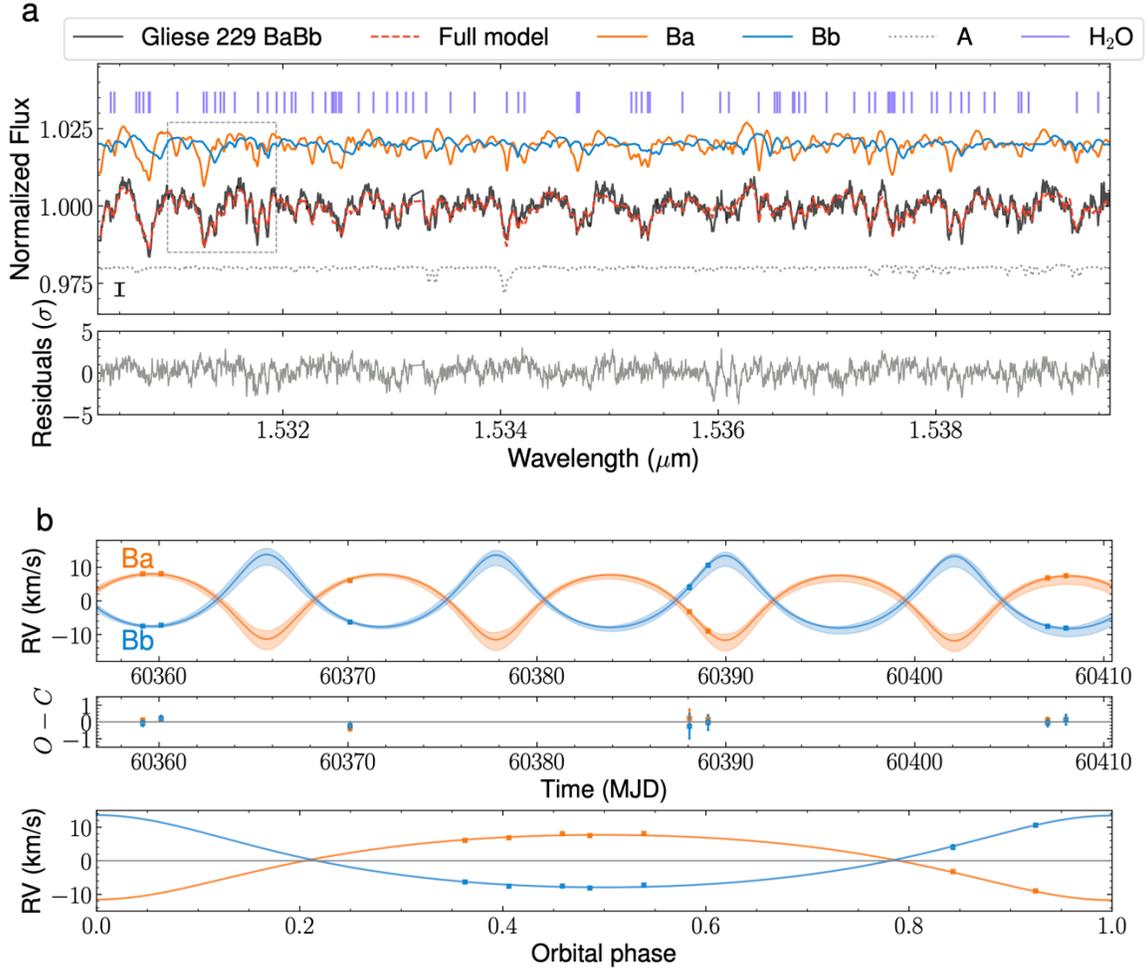

**Fig. 2 | CRIRES+ spectrum and spectroscopic orbit of Gliese 229 Ba and Bb.**
(a) A segment of the CRIRES+ spectrum from 2024-03-20 (black) used to compute radial velocities of Gliese 229 Ba and Bb. The region is dominated by water absorption lines from the brown dwarfs, whose positions are marked in purple. The orange and blue curves are spectral models for Ba and Bb, while the dashed gray curve is the CRIRES+ spectra of Gliese 229 A used to model stellar contamination. The three model components have been offset for clarity. The full model is shown in red. The median uncertainty of the spectrum is denoted by the 1σ error bar on the lower left. In the gray box, we highlight a region where distinct lines from Ba and Bb can be identified by eye. (b) The orange and blue points show the radial velocities of Gliese 229 Ba and Bb extracted from seven epochs of VLT/CRIRES+ spectra. Solid lines denote the joint CRIRES+ and GRAVITY orbit fit with 2σ uncertainty regions in shade. The middle panel shows the residuals of the best fit, and the bottom panel shows the phase-folded radial velocity orbit.

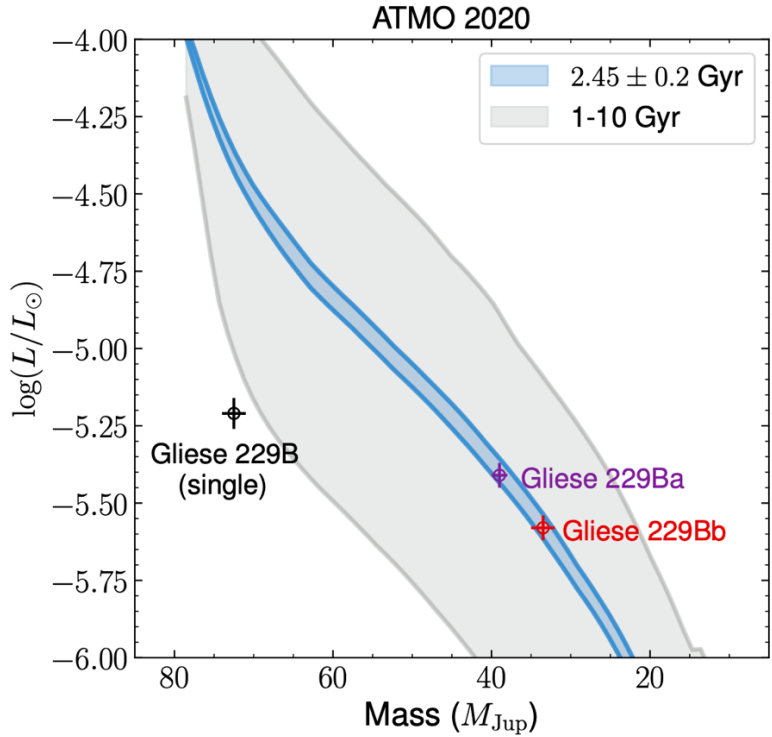

**Fig. 3 | Dynamical masses and inferred luminosities of Gliese 229 Ba and Bb from to ATMO 2020.**

The dynamical masses from our orbit fit and inferred luminosities of Gliese 229 Ba (purple) and Bb (red) from the ATMO 2020 evolutionary model. As a single brown dwarf, Gliese 229 B is under-luminous compared to model predictions for all plausible ages of the system from ref. 13. The mass-luminosity tension is also present for other models (see Extended Data Fig. 1). As a binary brown dwarf, the system is well-explained by the ATMO 2020 model for an age of 2.45±0.20 Gyr, resolving the mass-luminosity tension.

## Methods

**VLTI/GRAVITY observations and data reduction.** We observed Gliese 229 BaBb with GRAVITY[40] at the VLTI using the four Unit Telescopes of Paranal (program IDs: 0112.C-2369(A) and 2112.D-5036(A); PI: Xuan). We obtained data at five epochs: 26 and 30 December 2023, 28 February 2024, 29 March 2024, and 29 April 2024 (universal time). We used GRAVITY in the wide-angle dual-field mode[17], recently commissioned as part of the GRAVITY+ upgrade[41]. In this mode, the field is divided into two at the telescope level and carried independently within the GRAVITY delay lines. One field, centered on the star Gliese 229 A, is used by the GRAVITY fringe tracker[42,43] to stabilize the fringes by compensating for the atmospheric piston and vibrations in the system. The other field, centered on the companion (now known to be binary), is observed by the GRAVITY spectrometer. The scientific observations were conducted with medium spectral resolution (R=500) in the unpolarized mode. A log of the observations is given in Extended Data Table 1.

The reduction of the raw data was performed using the ESO GRAVITY pipeline v1.6.4 (ref. 44). This version of the pipeline can reduce the wide-angle data, but we had to disable the acquisition camera reduction to do so. In wide-angle mode, we could not use the fringe tracker to reference the phase (as is traditionally done for exoplanet observations; see ref. 45), but we could use the closure phase to detect the companion. The closure phases are averaged for each exposure, yielding several values per night.

The best datasets were obtained during the first two nights (see Extended Data Table 1). In December 2023, the two epochs showed closure phase values on the order of 40 degrees between 2 and 2.2 μm, with an SNR above 10. At longer wavelengths, $CH_4$ absorption and lower instrumental throughout prevent us from recording a robust closure phase. The closure phase signal was clear enough to confirm the binary nature of Gliese 229 BaBb. Moreover, injection-recovery tests show that the first epoch GRAVITY data are sensitive to objects 2-3 magnitudes fainter than Gliese 229 Ba at separations from 3 to 19 milliarcsecond, largely ruling out a third brown dwarf in the field. While the binary detection was clear, the data were too sparse to determine the orbital parameters, so we requested ESO Director's Discretionary Time to continue monitoring the object from February to late April, after which the target was no longer observable. The data quality was poor in February due to seeing conditions, and in March due to an issue with the pointing of the GRAVITY fibers. Despite this, a few below-average quality datasets were salvageable. The last dataset, obtained on UT 2024 April 29, was of high quality, benefiting from a recent instrumental upgrade of the VLTI beam compressor differential delay lines. During this last run, an SNR close to 20 was achieved, providing a clear detection to finalize the astrometric orbit of the binary.

**VLT/CRIRES+ observations and data reduction.** We observed the Gliese 229 system with the upgraded CRIRES+ (refs. 18,46) mounted on VLT (program ID: 0112.C-2369(B); PI: Xuan). We obtained seven epochs of data on 19 and 20 February 2024, 1, 19, and 20 March 2024, 7 and 8 April 2024 (universal time, see log in Extended Data Table 1). The wavelength setting H1567 and 0.2 arcsecond slit width were used to cover $H_2O$ and $CH_4$ absorption lines from 1.47-1.78 μm and achieve a spectral resolution of R≈100,000. The observations are taken in adaptive optics mode. For each epoch, we first observe the A0V telluric standard star 10 Lep (which is at a similar airmass as Gliese 229) and the primary star Gliese 229 A, before offsetting the slit to the companion's location ≈4.4 arcsec away. The relative astrometry of the companion is determined using the orbit from ref. 1. The CRIRES+ slit was set perpendicular to the position angle of the companion to minimize the leakage of starlight into the slit. We used the standard ABBA nodding scheme for background removal.

We reduced the data with a customized open-source pipeline *excalibuhr*[47]. It follows the general calibration steps of the ESO's pipeline CR2RES, including dark and flat correction, spectral order tracing, slit curvature tracing and initial wavelength solution. We removed the sky background via nod subtraction and combined individual exposures at each nodding position. The 1D spectra were then extracted using the optimal extraction method[48]. We used the spectrum of the standard star 10 Lep as a proxy to remove the telluric transmission features. The wavelength regions contaminated by strong telluric lines (with transmission less than 70%) were masked in the following analyses. Using observations of the telluric standard star, we carried out an additional wavelength calibration against a telluric transmission model generated by ESO's sky model calculator SkyCalc[49,50]. This was achieved by applying a third-order polynomial to the initial wavelength solution in each order and optimizing the correlation between the observed spectrum of the telluric standard star and the template spectrum.

On average, we achieved a S/N≈30 per wavelength channel per epoch at 1.57 μm for the extracted spectra of Gliese 229 BaBb, which includes emission from the companion and stellar contamination at the companion's location. To estimate the spectral resolution of our observations, we used the ESO sky software Molecfit[51] to fit the spectra of the telluric standard star. We find stable line spread functions across different nights with Gaussian profile widths of 3.05, 3.12, 3.28, 3.27. 3.05, 3.28, and 3.05 pixels, for seven epochs respectively. They correspond to an average resolving power of ≈100,000 as expected.

**Extraction of radial velocities from CRIRES+.** To calculate the RVs of Gliese 229 Ba and Bb, we fit the CRIRES+ spectrum from 1.510–1.583 μm, which covers two spectral orders. Each order is broken up into three chunks that are recorded on different detectors. The data from 1.45-1.50 μm are omitted due to significant telluric contamination. We also omit the data longward of 1.60 μm for two reasons. First, Gliese 229 Ba and Bb are extremely faint from 1.6 to 1.78 μm due to $CH_4$ absorption (see low-resolution spectrum in ref. 52), which results in lower S/N data.

Second, our preliminary fits show that the models provide a poorer match to the data beyond 1.6 μm. While we are using the most accurate $CH_4$ line list from ref. 53, ref. 54 showed that even this line list can produce discrepant $v\sin i$, radial velocity, and $T_{eff}$ measurements by fitting the spectrum of an isolated T dwarf. To avoid biasing the RV measurements, we focus on the water-dominated region from 1.510–1.583 μm, where the $H_2O$ line list from ref. 55 is shown to be accurate[54].

In the spectrum of Gliese 229 BaBb, we noticed atomic lines from Gliese 229 A, indicating a modest amount of stellar contamination from the bright host star (10 magnitudes brighter in H band). Therefore, we model the spectrum of Gliese 229 BaBb with three components: two brown dwarfs (Ba and Bb) and the primary star. The models for the brown dwarfs are generated using the temperature and abundances profiles from Sonora Elf Owl[20]. As the resolution limit of Elf Owl is R=5000, we re-compute the models at R=1,000,000 using the open-source radiative transfer code petitRADTRANS[56]. We include the line opacities of $CH_4$, $H_2O$, CO, $CO_2$, $H_2S$, $NH_3$, $PH_3$, $C_2H_2$, HCN, Na, K, and FeH, in addition to $H_2$-$H_2$, and $H_2$-He continuum opacities. To account for stellar contamination, we use the CRIRES+ spectrum of the star taken right before the Gliese 229 B exposures. Before fitting, we continuum-normalize each order of the Gliese 229 BaBb spectrum with a median filter of width 100 pixels (≈5 Å).

We fit the RV shift of the brown dwarfs at each observing date, the $v\sin i$ for each brown dwarf, flux scaling factors, and multiplicative error inflation terms. A different flux scaling factor is used for Gliese 229 Ba, Bb, and the primary star. To reduce the dimensionality, we optimize the linear flux scaling terms and error inflation terms at each iteration following ref. 57. In the fit, we rotationally broaden the atmospheric models using the code from ref. 58, apply the RV shifts, and convolve the models to R=100,000 with a Gaussian profile. Next, we apply the optimal scale factors to the respective models to construct the combined model (Fig. 2a) and apply the same median filter to the combined model. The posterior is sampled with the nested sampling code DYNESTY[59], and we use 1000 live points. We find that on average, the host star contributes ≈20% of the total flux in the Gliese 229 BaBb spectrum. Because the lines from the M1V primary star are very distinct from T dwarf lines (e.g. Fig. 1), they do not impact our RV measurements.

After obtaining the RV posteriors, we apply barycentric corrections for each night using tools in the Astropy package[60] and subtract the RV of the primary star taken from ref. 61. The resulting RV points of Gliese 229 Ba and Bb are shown in Fig. 2b, and included in Extended Data Table 3. The statistical errors on the measured RVs are typically ≈0.1 km/s. We consider several sources of systematic uncertainties. First, we measure the stellar RVs over the same nights to assess the instrumental jitter. The procedure is described below, and adds a 0.1 km/s uncertainty. Second, we consider the impact that uncertain atmospheric parameters have on the retrieved RVs by repeating the spectral fits with a range of different models. Besides the fiducial model (850 K +

750 K), we consider the following $T_{eff}$ combinations: (900 K + 800 K), (900 K + 750 K), (850 K + 700 K), (850 K + 800 K), (800 K + 750 K). We set log($g$)=5.0, C/O=0.68, and [M/H]=0.0 for all models. The abundances are chosen to match those of the host star, which has a nearly solar metallicity[62,63] and C/O=0.68±0.12 (ref. 64). The log($g$) is fixed because the evolutionary models predict a relatively small range of variation in log($g$) (see Extended Data Table 2). In addition, we fix the vertical eddy diffusion parameter log($K_{zz}$) to 2.0 as found by ref. 20. We use the scatter in RV values derived from each fit as an independent source of systematic error. These add systematic uncertainties on the order of ≈0.2-0.7 km/s, depending on the epoch.

We compute cross-correlations functions (CCF) of the primary star spectra to verify the stability of the CRIRES+ wavelength solution and line spread function (Extended Data Fig. 3). We adopt a PHOENIX-ACES model[65] with $T_{eff}$=3800 K and log($g$)=4.5 for the primary star. Over the 2.5 months observing period, the RV change caused by orbital motion of Gliese 229 A around the system barycenter is <2 m/s, which we ignore. Approximating the stellar CCFs as Gaussian functions, we measure the stellar RVs as the center of the Gaussian. We find that the stellar RV is stable at the 0.1 km/s level across the seven observing epochs.

From the CRIRES+ fits, we find that the two brown dwarfs have projected rotation rates ($v$sin$i$) below our measurement limit. The 3σ upper limits of $v$sin$i$ for Gliese 229 Ba and Bb are <0.6 km/s and <0.7 km/s, respectively. If the two brown dwarfs are tidally synchronized, their rotational velocities would be ≈0.4 km/s. Assuming their rotational axes are aligned with the orbital axis, this implies $v$sin$i$≈0.2 km/s, which is well below the size of the line spread function (≈3 km/s) for CRIRES+. Thus, our non-detection of spin is consistent with the brown dwarfs being tidally locked, or nearly tidally locked, which is expected based on their tidal despinning time (see the 'Dynamics' section).

**Bulk properties of Gliese 229 Ba and Bb.** Using ATMO 2020 evolutionary models[9], we estimate the component masses of Gliese 229 Ba and Bb that best reproduce the bolometric luminosity of log($L/L_\odot$) of -5.21±0.05 (ref. 7) and GRAVITY K band flux ratio of 0.47±0.03. We additionally include J, H, K magnitudes of the combined source[66] as constraints in our fit. ATMO 2020 includes three separate models with differing amounts of non-equilibrium (NEQ) chemistry. We adopt the "NEQ weak" model but note that the results are similar if we used "NEQ strong" or "CEQ". We use ATMO 2020 tables with pre-computed Mauna Kea Observatories (MKO) magnitudes. While the GRAVITY K band transmission profile is not identical to that of MKO K, the flux ratio measurement effectively divides out the transmission function. Our fit is parameterized with three parameters: mass ratio, age, and total mass. We place a Gaussian prior of 72.5±1.3 $M_{Jup}$ on total mass, as derived from our orbit fit. For a given set of masses and age, we interpolate to obtain the log($L/L_\odot$), and J, H, K magnitude of each brown dwarf, requiring that their combined magnitudes and luminosities match the observed values. We sample the posterior using a Markov Chain Monte Carlo method[67] with 10000 steps and 30 walkers. The first 2000 steps are discarded as burn-in. Overall, the ATMO 2020 models match the observations well for an age of 2.45±0.2 Gyr (Fig. 3). The inferred age is somewhat model-dependent, but we find that ages of 2-4 Gyr generally match the properties of the binary brown dwarf by considering alternative evolutionary models in Extended Data Fig. 1.

From the ATMO 2020 model, we also interpolate for the $T_{eff}$, $\log(g)$, and $\log(L/L_\odot)$ of each brown dwarf, which we tabulate in Extended Data Table 2. We adopt the closest grid points in Sonora Elf Owl to these values to compute high-resolution spectral models and fit the CRIRES+ spectra. We emphasize that these $T_{eff}$ estimates are model-based. Upcoming JWST spectroscopy of Gliese 229 BaBb from 1-15 μm (GO3762; PI: Xuan) should enable robust two-component spectral fits and provide independent estimates of the bulk properties for each brown dwarf.

We perform a second estimate of the bulk properties of Gliese 229 Ba and Bb using calibrated empirical relations for field brown dwarfs from ref. 68. First, we estimate individual absolute MKO $M_K$ magnitudes of Gliese 229 Ba and Bb from their combined-light MKO $K$ magnitude ($K_{MKO} = 14.36 \pm 0.05$)[66], the GRAVITY $K$ band flux ratio (0.47 ± 0.03), and the system parallax (173.574 ± 0.017 mas)[14]. This yields $M_{K,Ba} = 15.98 \pm 0.05$ mag and $M_{K,Bb} = 16.80 \pm 0.05$ mag. Next, using the MKO $M_K$-Lbol and MKO $M_K$-$T_{eff}$ relations for field objects in ref. 68, we find $\log(L/L_\odot)$, Ba = -5.36 ± 0.07 dex, $T_{eff, Ba}$ = 810 ± 55 K, and $\log(L/L_\odot)$, Bb = -5.56 ± 0.07 dex, $T_{eff, Bb}$ = 694 ± 55 K. These values are consistent with those inferred from the ATMO 2020 evolutionary models at the ≈1σ level. The closest matching spectral types are T7 for Gliese 229 Ba and T8 for Gliese 229 Bb.

**Orbit fits.** To derive orbital parameters, we jointly fit the GRAVITY closure phases and CRIRES+ RVs. Instead of computing positions from the closure phases for each epoch, we directly model them in the orbit fit. Not only does this take into account the multiple possible positions at each epoch, it also avoids intermediate products, preserving noise properties. We implemented this joint model in two different frameworks: a frequentist approach in PMOIRED[22] and a Bayesian approach in Octofitter[23]. These methods independently arrive at consistent results. The methods were additionally validated using high quality GRAVITY data and high S/N radial velocities from VLT/UVES[69] for a binary star system, where we confirm that a close-phase orbital fit and an orbit fit using per-epoch separations and position angles yielded the same result. In both codes, we adopt a standard coordinate system for the orbit where +X points East, +Y points North, and +Z points away from the observer.

For the PMOIRED analysis, the best orbit is found by gradient descent, first on the radial velocity data and then on the joint model after adding the closure phase data. We only include the GRAVITY data from 2.05-2.18 μm as strong methane absorption results in extremely low S/N past 2.18 μm. In addition, the wavelength channels are binned to six points over the 2.05-2.18 μm range. In order to better estimate the final uncertainties, bootstrapping is used: 5000 random datasets are generated using sampling with replacement, and each time an orbital solution is fitted from a first guess drawn around the best values with four times the uncertainties. Bootstrapping has been shown to mitigate the correlations in interferometric data analysis[70]. GRAVITY data are correlated, primarily because closure phases share baselines and baselines share telescopes (as formalized in ref. 71). Moreover, data taken at the same time and with the same telescope triples have experienced the same biases from atmospheric turbulence and same calibration processes. To account for these correlations, all closure phases from the same date and baseline triangle are drawn together on the bootstrapping. First, we search for the best-fit orbit to the radial velocity data alone. This leads to an excellent solution with P=12.12±0.04 d, e=0.22±0.03, q=0.91±0.03 and a reduced $\chi^2$ of 1.3. We allow for a RV offset term, $\gamma_{RV}$ to account for possible inaccuracies in the systematic RV of the system. Next, we perform a joint fit to the GRAVITY and CRIRES+ data. The results are shown in Table 1, and the relative orbit of the binary brown dwarf is plotted in Extended Data Fig. 6. We adopt the PMOIRED results as the baseline values in this paper.

For the Octofitter analysis, we completed joint Bayesian modeling of both the CRIRES+RVs and GRAVITY data. We used non-reversible parallel tempering[72-75] to search the entire multi-modal parameter space globally for the best-fitting parameter values. Rather than working with the closure phases directly, this analysis first converts the closure phases into a set of non-redundant kernel phases for each wavelength[76]. This improves the accuracy of the model uncertainties compared to working directly with the closure phases, which share baselines (mitigated in the PMOIRED analysis after-the-fact using bootstrapping). Finally, we add an additional kernel-phase "jitter" term for each epoch of data (five in total). This term allows the model to absorb some amount of systematic calibration error in the GRAVITY data, again resulting in more realistic uncertainties in the final model parameters. For this model, we included data in the 2.025 - 2.15 μm range with no spectral binning. The orbital parameters from the joint model are listed in Table 1, and are consistent with PMOIRED results at the ≈1.5 sigma level. We find strong evidence that, when combining the GRAVITY data with the CRIRES+ RVs, the orbit solution is uniquely determined and no secondary modes in the posterior are significant.

We provide posterior predictions of the relative separation and position angle of Gliese 229 Ba-Bb in Extended Data Table 4. We stress that these are inferred values and not statistically independent like traditional astrometry, as they are derived from a joint analysis of all epochs. They should not be used as inputs to an orbit fit, as they themselves are the outputs of such a fit. Instead, orbit fits should use the GRAVITY closure phases.

**Dynamics.** Given their small separation, Gliese 229 Ba and Bb are probably tidally locked with each other, with rotation periods equal to the orbital period of 12 days. We quantify the tidal locking timescale using ref. 77. With an initial spin velocity of 20 km s−1 and initial radius of 1 $R$Jup, we find a despinning time of about 2 Gyr, which is shorter than or comparable with the estimated system age of approximately 2–6 Gyr (ref. 13). As noted earlier, our CRIRES+ analysis shows that both brown dwarfs have $v\sin i$ < 0.7 km s$^{-1}$, which is consistent with them being tidally locked.

In the current configuration of Gliese 229 A-BaBb, the highly eccentric and misaligned outer AB orbit ($e \approx 0.85$)[1] could induce secular perturbations that pump up the eccentricity of the inner BaBb orbit by means of the eccentric von Zeipel–Lidov–Kozai mechanism[78]. Consequently, tidal interactions may shrink the BaBb orbit. We estimate that the Kozai secular precession timescale given by equation (3) in ref. 79 is about 0.2 Myr for Gliese 229 Bb. The secondary brown dwarf also undergoes precession from the quadrupole potential from its tidal and rotational bulges and from the leading order effects of general relativity. If these effects operate on a shorter timescale, they could suppress Kozai oscillations. For our brown dwarfs, we adopt tidal parameters $Q = 3 \times 10^4$, $k2 = 0.565$ based on Jupiter. The exact values of $Q$ and $k2$ are unknown for brown dwarfs but estimates from hot Jupiters generally produce values within one order of magnitude of Jupiter's values[80]. We estimate the precession rates using equations (6)–(8) in ref. 79 and find that the precession rate owing to general relativity is the fastest, with a corresponding timescale of about 0.6 Myr, which is still longer than the Kozai timescale. In the absence of further perturbations or bodies, the triple system may therefore undergo Kozai oscillations. However, detailed $N$-body simulations and follow-up work are required to further evaluate the dynamical state of the system.

**Data availability** The reduced CRIRES+ and GRAVITY data will be made public through Zenodo[81] at: https://doi.org/10.5281/zenodo.13851639.

**Code availability** The CRIRES+ data reduction was performed with *excalibuhr*. (https://github.com/yapenzhang/excalibuhr). The orbit fits were performed with PMOIRED (https://github.com/amerand/PMOIRED) and Octofitter (https://sefffal.github.io/Octofitter.jl/dev/). The atmospheric models were generated using inputs from Sonora Elf Owl https://zenodo.org/records/10385821 and the petitRADTRANS radiative transfer tool available at https://petitradtrans.readthedocs.io/. ATMO 2020 models are available for download at http://opendata.erc-atmo.eu.

**Acknowledgements** J.W.X. thanks S. Mukherjee, T. Dupuy, B. Lacy, C. Morley, J. Fortney, Y. Zhou, G. Hallinan, and S. Kulkarni for helpful discussions. J.W.X. acknowledges support for this work through the NASA FINESST Fellowship award 80NSSC23K1434. S.L. acknowledges the support of the French "Agence Nationale de la Recherche" (ANR), under grants ANR-21-CE31-0017 (project ExoVLTI) and ANR-22-EXOR-0005 (PEPR Origins). M.C. Liu acknowledges support from the Gordon and Betty Moore Foundation through grant GBMF8550. J.Wa. acknowledges support from NASA Grant 80NSSC23K0280. P.K. acknowledges funding from the European Research Council (ERC) under the European Union's Horizon 2020 research and innovation program (project UniverScale, grant agreement 951549). P.G. acknowledges the financial support provided by FCT/Portugal through grants PTDC/FIS-AST/7002/2020 and UIDB/00099/2020. This paper is based on observations collected at the European Southern Observatory under ESO programme 0112.C-2369(A), 0112.C-2369(B), and 2112.D-5036(A). This research has made use of the Jean-Marie Mariotti Center Aspro service available at http://www.jmmc.fr/aspro. This work has benefitted from The UltracoolSheet at http://bit.ly/UltracoolSheet, maintained by Will Best, Trent Dupuy, Michael Liu, Aniket Sanghi, Rob Siverd, and Zhoujian Zhang, and developed from compilations by ref 68 and ref 82-88.

**Author Contributions** J.W.X. led the telescope proposals for Gliese 229, performed the CRIRES+ spectral fitting to obtain radial velocities, and wrote the manuscript. A.M., W.T., and D.B. led the closure phase modeling and orbital analysis. Y.Z. reduced the raw CRIRES+ data. S.L. reduced the raw GRAVITY data. D.M., R.O., M.L., and A.B. provided advice on the writing and figures. J.K. computed the GRAVITY sensitivity limits. K.B. provided guidance on the system dynamics. A.Sa. performed the analysis with empirical relations. H.K. provided advice on the CRIRES+ spectral fits. R.B.-R and M.Sa. obtained confirmation Keck/NIRC2 images. The remaining authors constitute the GRAVITY team and commented on the manuscript.

**Competing interests** The authors declare no competing interests.

**Correspondence and requests for materials** should be addressed to Jerry W. Xuan.

**Reprints and permissions information** is available at [www.nature.com/reprints](www.nature.com/reprints)

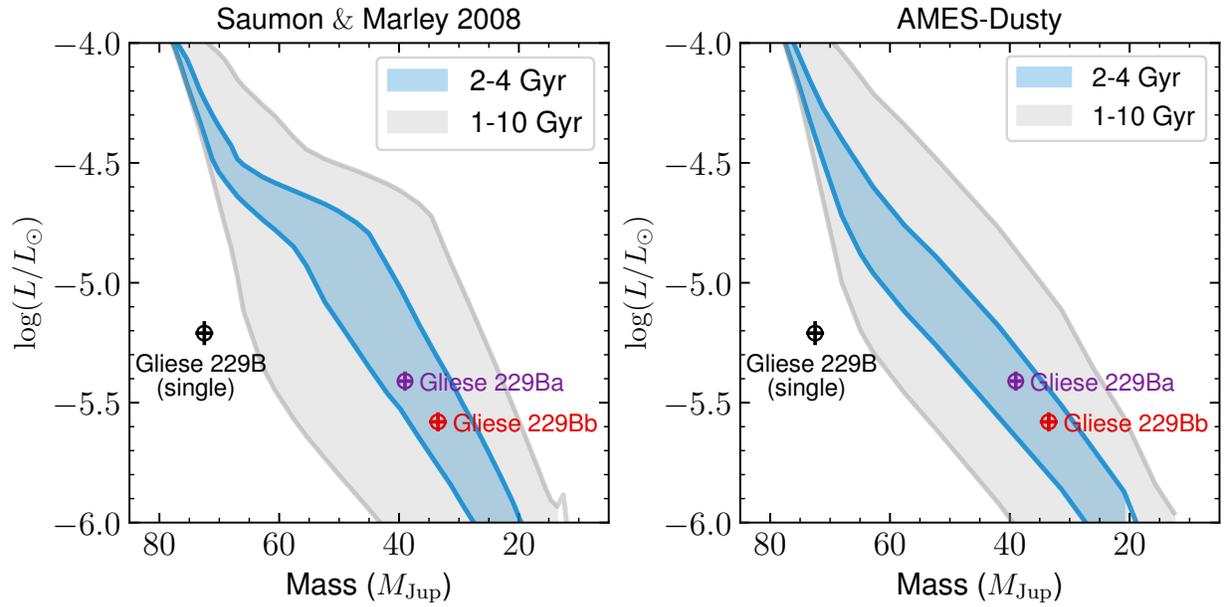

**Extended Data Fig. 1 | Dynamical masses and inferred luminosities of Gliese 229 Ba and Bb compared to Saumon & Marley 2008 and AMES-Dusty models.**
The dynamical masses and estimated luminosities of Gliese 229 Ba (purple) and Bb (red) from the ATMO 2020 evolutionary model fit. As a single brown dwarf, Gliese 229 B is underluminous compared to model predictions even at 10 Gyr (leftmost gray line). As a binary brown dwarf, the system is consistent with the Saumon & Marley 2008[8] and AMES-Dusty[89] models for an age of ≈2-4 Gyr.

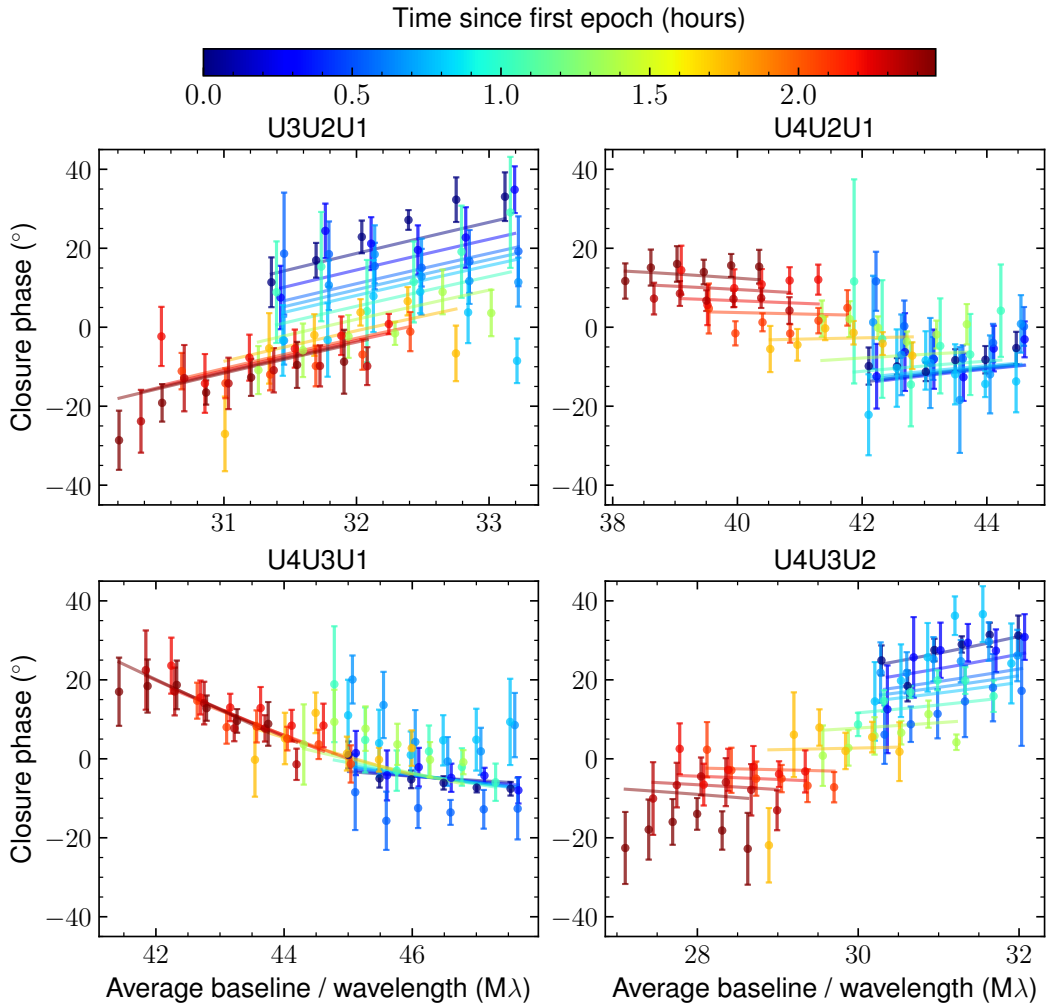

**Extended Data Fig. 2 | GRAVITY closure phase measurements of Gliese 229 BaBb on 2023-12-26.**

The GRAVITY closure phases measurements in the first epoch (2023-12-26). The data are in points, and the models are shown as lines. Each panel is for a different baseline triangle between the four unit telescopes at the Very Large Telescope (U1, U2, U3, U4). The color code indicates the time since the first data point (in hours). The data are well described by the model, with a majority of residuals at the <2σ level. A single source would have zero closure phases throughout.

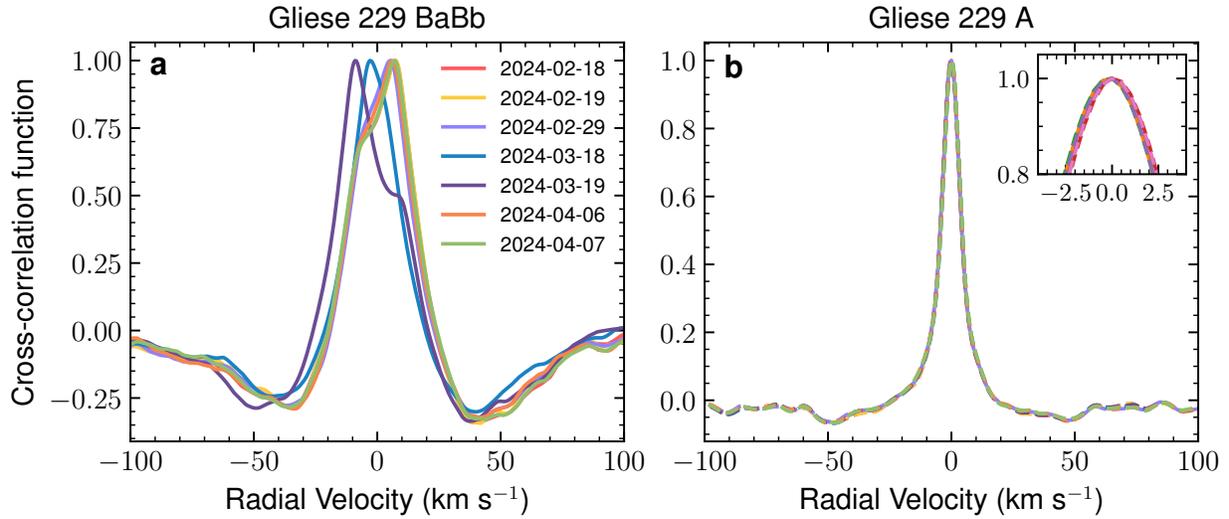

**Extended Data Fig. 3 | Cross-correlation functions of the CRIRES+ spectra of Gliese 229 B and A.**

(a) Cross-correlation functions (CCFs) between CRIRES+ spectra of Gliese 229 B and an atmospheric model with $T_{\rm eff}$=900 K, $\log(g)$=5.0 computed using Sonora Elf Owl temperature and chemistry profiles. The CCF shapes are distorted and variable over time, characteristic of a double-lined spectroscopic binary. (b) CCFs between CRIRES+ spectra of Gliese 229 A and a PHOENIX-ACES model[65] with $T_{\rm eff}$=3800 K and $\log(g)$=4.5. The inset shows a zoom-in of the CCF peak. The stellar RVs are stable at the 0.1 km/s level over the observing period, validating the wavelength solution of CRIRES+.

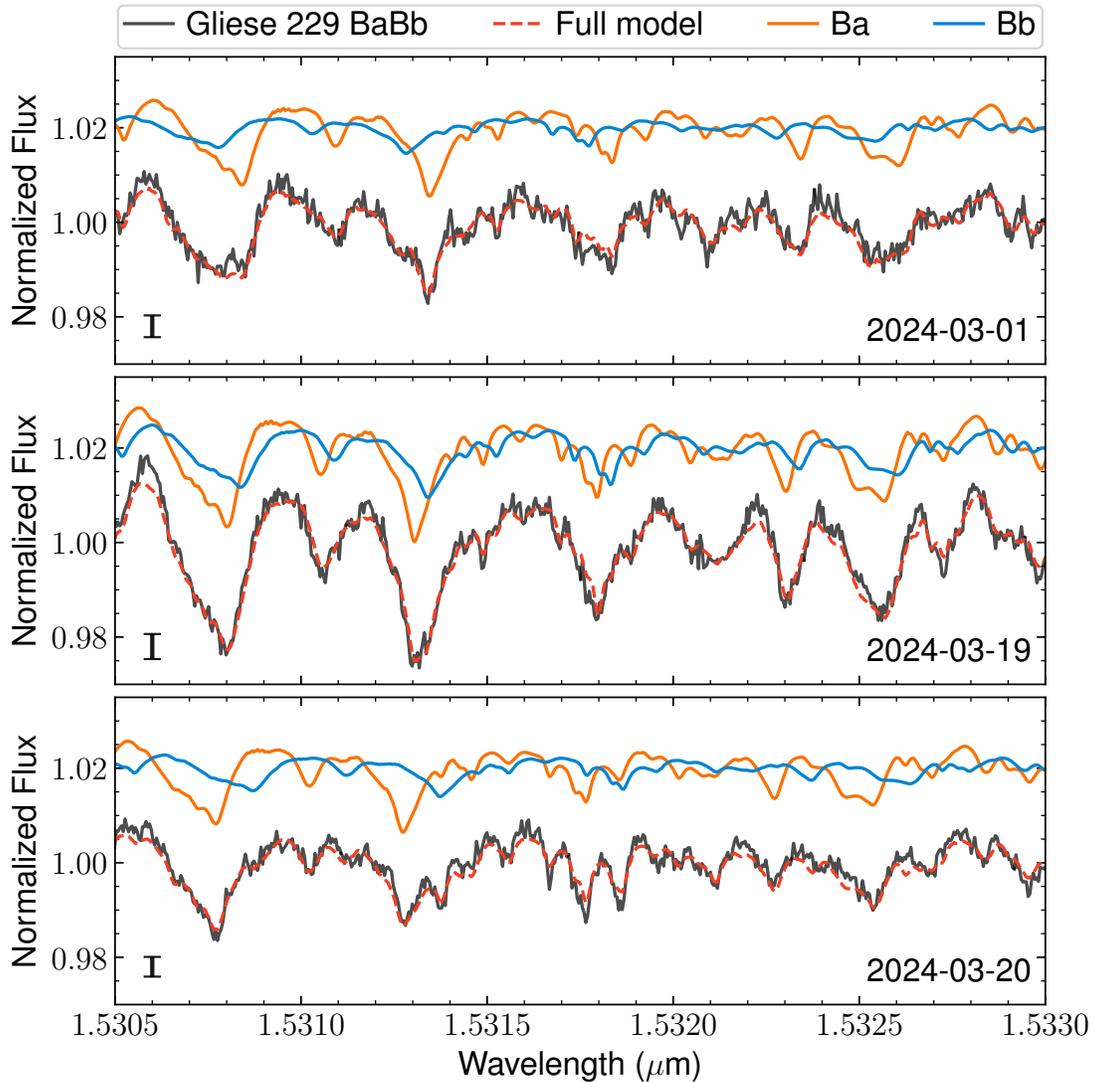

**Extended Data Fig. 4 | A zoom-in of the CRIRES+ spectra of Gliese 229 B on three different nights.**
A small portion of the CRIRES+ spectra on three different nights where we achieve the highest S/N (black). The Ba and Bb models are shown in teal and purple, while the full model is in red. The median uncertainties for the spectra are denoted as error bars on the lower left (1σ). Absorption lines from the two brown dwarfs can be seen combining over the observing sequence. The data from 2024-03-19 were taken with the best seeing conditions and consequently contains the highest flux from the brown dwarfs and minimal stellar contamination from Gliese 229 A. Therefore, the lines appear deeper for this epoch.

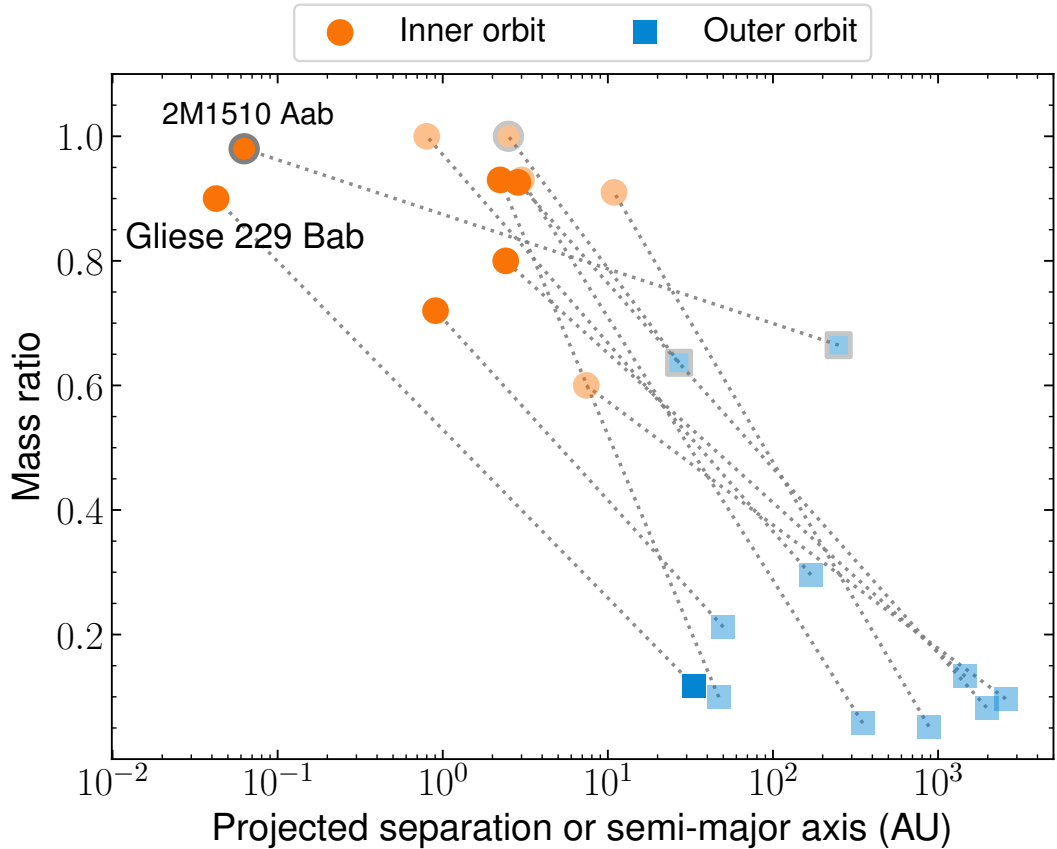

**Extended Data Fig. 5 | Gliese 229 A-Bab and other brown dwarf binaries in triple systems.** For each system, we show the orbital separation and mass ratio of the brown dwarf binary in orange. The separation of the outer orbit (i.e. that between the brown dwarfs and the third component) and mass ratio of the brown dwarf binary relative to the total system mass is in blue. Many systems have prohibitively long orbital periods or lack published orbit solutions; we use transparent points to denote projected separations, and opaque points for semi-major axes. Each system is connected with a gray dotted line. We label the similarly tight binary 2M1510 Aab from ref. 30. The circles with gray outlines are triple brown dwarf systems, where all components are substellar. Among brown dwarf binaries orbiting stars, Gliese 229 Bab has an inner orbit more than an order of magnitude smaller than other systems. The parameters for other systems are taken from refs. 84, 90-96.

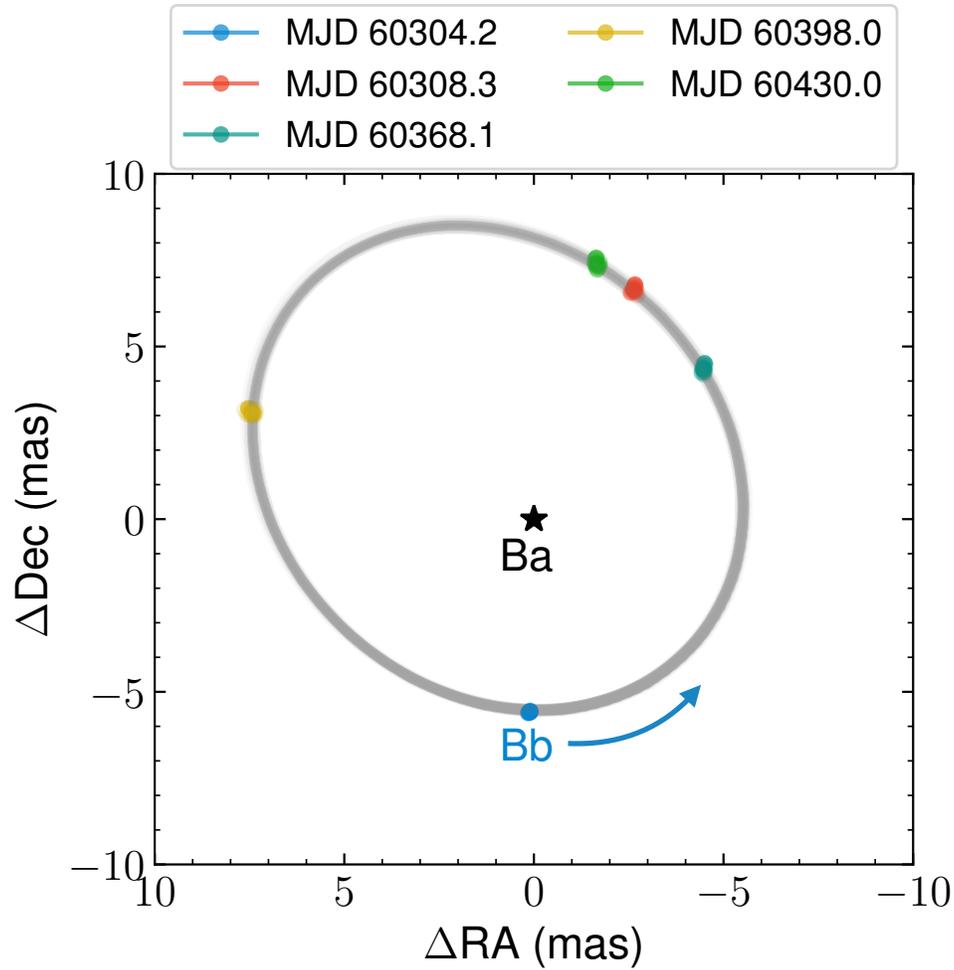

**Extended Data Fig. 6 | Relative orbit of Gliese 229 Bb with respect to Gliese 229 Ba from GRAVITY and CRIRES+.**

Random draws of the relative astrometric orbit of Gliese 229 Ba-Bb from the PMOIRED fit are shown as gray curves. The position of the primary brown dwarf Gliese 229 Ba is marked with a star at the origin. The colored points show random draws of the predicted astrometric positions of Bb with respect to Ba from the joint GRAVITY and CRIRES+ orbit fit over the five observing epochs.

**Extended Data Table 1 | GRAVITY wide and CRIRES+ observation log for Gliese 229 BaBb**

| Instrument | DATE START (UT) | DATE END | NEXP/NDIT/DIT(seconds) | AIRMASS | TAU0 (milli-second) | SEEING (arcsec) |
|---|---|---|---|---|---|---|
| GRAVITY Wide | 2023-12-26T04:14:07 | 2023-12-26T06:48:36 | 13/4/100 | 1-1.16 | 5-20 | 0.5-1.1 |
| GRAVITY Wide | 2023-12-30T05:47:25 | 2023-12-30T06:32:50 | 4/4/100 | 1.1-1.2 | 5-10 | 0.4-0.8 |
| GRAVITY Wide | 2024-02-28T02:57:39 | 2024-02-28T03:36:49 | 3/4/100 | 1.2-1.4 | 5-6 | 0.7-0.9 |
| GRAVITY Wide | 2024-03-29T00:31:24 | 2024-03-29T01:17:35 | 4/4/100 | 1.1-1.3 | 8-12 | 0.4-0.6 |
| GRAVITY Wide | 2024-04-29T23:15:22 | 2024-04-29T23:29:21 | 2/4/100 | 1.3-1.4 | 4-10 | 0.7-0.8 |
| CRIRES+ | 2024-02-19T02:27:37 | 2024-02-19T04:10:34 | 6/1/900 | 1.1-1.4 | 7-18 | 0.6-1.15 |
| CRIRES+ | 2024-02-20T02:22:28 | 2024-02-20T03:00:26 | 2/1/900 | 1.1 | 5-7 | 0.8-1.5 |
| CRIRES+ | 2024-03-01T01:45:04 | 2024-03-01T03:28:13 | 6/1/900 | 1.1-1.4 | 7-9 | 0.5-0.8 |
| CRIRES+ | 2024-03-19T00:54:23 | 2024-03-19T02:37:11 | 6/1/900 | 1.1-1.5 | 6-8 | 0.4-0.7 |
| CRIRES+ | 2024-03-20T00:36:04 | 2024-03-20T02:18:21 | 6/1/900 | 1.1-1.4 | 2-6 | 0.6-0.7 |
| CRIRES+ | 2024-04-07T00:12:23 | 2024-04-07T01:12:00 | 4/1/600 | 1.2-1.4 | 4-6 | 0.6-0.9 |
| CRIRES+ | 2024-04-08T23:31:31 | 2024-04-08T00:24:28 | 4/1/600 | 1.1-1.2 | 7-8 | 0.7-1.0 |

**Extended Data Table 2 | Bulk properties of Gliese 229 BaBb inferred from the ATMO 2020 evolutionary model.**

| Parameter | ATMO 2020 |
| --- | --- |
| Mass ratio | 0.87±0.03 |
| Age (Gyr) | 2.45±0.20 |
| $T_{eff,Ba}$ (K) | 860±20 |
| $T_{eff,Ba}$ (K) | 770±20 |
| $\log(g)_{Ba}$ (dex) | 5.11±0.01 |
| $\log(g)_{Bb}$ (dex) | 5.03±0.01 |
| $\log(L/L_\odot)_{Ba}$ | -5.41±0.04 |
| $\log(L/L_\odot)_{Bb}$ | -5.58±0.04 |

**Extended Data Table 3 | Radial velocities of Gliese 229 Ba and Bb from VLT/CRIRES+.**

| Time (MJD) | RV$_{Ba}$ (km/s) | RV$_{Bb}$ (km/s) |
|---|---|---|
| 60359.14 | 8.07±0.12 | -7.51±0.19 |
| 60360.11 | 8.13±0.14 | -7.22±0.19 |
| 60370.11 | 6.07±0.17 | -6.31±0.25 |
| 60388.07 | -3.21±0.54 | 4.09±0.75 |
| 60389.058 | -8.97±0.23 | 10.60±0.46 |
| 60407.03 | 6.85±0.16 | -7.55±0.24 |
| 60408.00 | 7.49±0.27 | -8.10±0.32 |

**Extended Data Table 4 | Derived relative astrometry of Gliese 229 Ba-Bb from the Octofitter fit.**

| Date [UTC] | Projected separation [mas] | Position Angle [deg] | Separation [AU] |
| --- | --- | --- | --- |
| 2023-12-26 05:15:42 | 5.58± 0.01 | 178.4 ± 0.9 | 0.0338 ± 0.0001 |
| 2023-12-30 06:11:23 | 7.10±0.05 | -22.0± 0.1 | 0.0453±0.0002 |
| 2024-02-28 03:13:29 | 6.19±0.04 | -46.5± 0.3 | 0.0413±0.0001 |
| 2024-03-29 00:31:14 | 7.95± 0.03 | 67.6 ±0.5 | 0.0488±0.0001 |
| 2024-04-29 23:15:12 | 7.49± 0.04 | -13.5± 0.2 | 0.0468±0.0002 |

As noted in Methods, these values should not be used directly in orbits fits.